\documentclass[prd,twocolumn,amsmath,aps,floats,amssymb, floatfix, superscriptaddress, nofootinbib]{revtex4}

\pdfoutput=1

\usepackage{graphicx}
\usepackage{wasysym}
\usepackage{amsfonts}
\usepackage{subfigure}
\usepackage{hyperref}
\usepackage{color}        
\usepackage{ulem}

\def\be{\begin{equation}}
\def\ee{\end{equation}}
\def\bea{\begin{eqnarray}}
\def\eea{\end{eqnarray}}
\newcommand{\vs}{\nonumber\\}

\newcommand{\ec}[1]{Eq.~(\ref{eq:#1})}
\newcommand{\Ec}[1]{(\ref{eq:#1})}
\newcommand{\eql}[1]{\label{eq:#1}}

\newcommand{\rf}[1]{\ref{fig:#1}}

\begin{document}

\title{Imprint of primordial non-Gaussianity on dark matter halo profiles}

\author{Azadeh Moradinezhad Dizgah}
\affiliation{Department of Physics, University at Buffalo, The State University of New York, Buffalo, NY 14260-1500, USA}

\author{Scott Dodelson}
\affiliation{Fermilab Center for Particle Astrophysics, Fermi National Accelerator Laboratory, Batavia, IL 60510-0500, USA}
\affiliation{Kavli Institute for Cosmological Physics, Enrico Fermi Institute, University of Chicago, Chicago, IL 60637, USA}
\affiliation{Department of Astronomy \& Astrophysics, University of Chicago, Chicago, IL 60637, USA}
\author{Antonio Riotto}
\affiliation{University of Geneva, Department of Theoretical Physics and Center for Astroparticle Physics (CAP)\\ 24 quai E. Ansermet, CH-1211 Geneva 4, Switzerland}

\begin{abstract}
\noindent
We study the impact of primordial non-Gaussianity on the density profile of dark matter halos by using  the semianalytical model introduced recently by Dalal {\it et al.} which relates the peaks of the initial linear density field to the final density profile of dark matter halos. Models with primordial non-Gaussianity typically produce an initial density field that differs from that produced in Gaussian models. We use the path-integral formulation of excursion set theory  to calculate the non-Gaussian corrections to the peak profile and derive the statistics of the peaks of the non-Gaussian density field. In the context of the semianalytic model for halo profiles, currently allowed values for primordial non-Gaussianity would increase the shapes of the inner dark matter profiles, but only at the sub-percent level except in the very innermost regions.
\end{abstract}

\maketitle

\section{Introduction}
Observations of both the cosmic microwave background (CMB) and large-scale structure (LSS) are compatible with adiabatic, nearly scale-invariant and Gaussian primordial perturbations in agreement with the predictions of the simplest inflationary models \cite{Lyth:1998xn,Ade:2013ydc}. Although successful in explaining the current observations, there are still open questions about the physics of inflation. Primordial non-Gaussianity is a sensitive probe of the interactions of quantum fields during inflation and hence contains important information about the fundamental physics responsible for inflation beyond that contained in the power spectrum \cite{Bartolo:2004if}.

The CMB provides a clean probe of primordial non-Gaussianity since the perturbations are still in the linear regime and unprocessed. Extracting information about primordial non-Gaussianity from large-scale structure is more complex since gravity couples Fourier modes to the point that imprints of the initial distribution are hidden. To constrain primordial non-Gaussianity (PNG) from large-scale structure, we need to identify a signature that can be  produced only by primordial non-Gaussianity and not by gravitational instability. One that has emerged in recent years is scale-dependent bias~\cite{Dalal:2007cu,LoVerde:2007ri,McDonald:2008sc,Giannantonio:2009ak,Desjacques:2010gz,Bartolo:2010ec,Shandera:2010ei,Paranjape:2011ak,Bruni:2011ta,Baldauf:2011bh,Jeong:2011as,Desjacques:2011jb,Musso:2012ch,Adshead:2012hs,D'Aloisio:2012hr,Desjacques:2013qx,Lidz:2013tra}. In this paper we investigate whether primordial non-Gaussianity leaves a distinctive imprint in a different arena, on the profile of dark matter halos.

In principle, if PNG led to a difference in the dark matter halo profile, we could detect this in a variety of ways. Direct mass maps are possible via weak lensing measurements; these maps though are unlikely to have the sensitivity to distinguish small changes in profiles. Another possibility is that a larger density in the interior of the halos would lead to a larger detection of dark matter annihilation in sensitive gamma-ray experiments. Although there are many astrophysical and particle physics uncertainties, the $\rho^2$ scaling of the signal (where $\rho $ is the density profile of the halo) could conceivably promote even small effects to prominence. Another area where the halo profiles assume importance is in the context of the halo model of large-scale structure, which is often used to model signals seen in galaxy surveys. One could imagine a scenario in which a change in the dark matter profile could propagate to the statistics observed in surveys if the changes were significant.

Within the standard $\Lambda \text{CDM}$ cosmology, halos form hierarchically through the mergers and accretion of previous generations of virialized objects. Therefore the formation of dark matter halos is a complex, nonlinear process. Nevertheless $N$-body simulations show regularity in properties of the final halos. In particular they indicate a universal density profile for the final halos, the so-called NFW profile \cite{Navarro:1995iw,Navarro:1996gj}, with $\rho \propto r^{-3}$ at large radii and $\rho \propto r^{-1}$ at small radii . The origin of this universal profile has been a long-standing question. In a recent paper \cite{Dalal:2010hy}, Dalal {\it et al.} proposed a semianalytical model to explain the origin of the NFW profile. They suggest that the profile of the collapsed halos can be determined in terms of their precursor peaks. Their model consists of two parts: calculating the averaged profile of a peak in the linear density field and a mapping between the properties of the initial peak to the properties of the final halo. Adiabatic contraction and dynamical friction are the main physical effects that transform the initial peak shape to the final halo profile.

Since in this model the density profile of the dark matter halos is determined in terms of the peaks of the linear density field, it allows us to trace the impact of primordial non-Gaussianities on the profile of the collapsed halos. We quantify the non-Gaussian corrections to the profile of the peaks of the linear density field using the path-integral formulation of excursion set theory recently developed by Maggiore and Riotto (MR)\cite{Maggiore:2009rv,Maggiore:2009rw,Maggiore:2009rx} and then propagate these changes to the shapes of the final dark matter halos. 

In principle one could study the effect of primordial non-Gaussianity on halo profiles by running simulations. Indeed, there has been some work in this direction~\cite{Smith:2010fh} which provided some fuel for the claim that non-Gaussianity does affect halo profiles.  This work is complimentary: because we attack the problem semianalytically, we can probe the profile much closer to the center without any resolution issues. Of course, the downside is the possibility that the method of Ref.~\cite{Dalal:2010hy} does not capture all of the physics relevant to this problem.

This paper is organized as follows. In Sec. \ref{sec:halo_model}, we review the semianalytical model of Dalal {\it et al.} to explain the origin of the NFW profile. In Sec. \ref{sec:path_integral} we review the path-integral formulation of excursion set theory and in Sec. \ref{sec:con_prob} we apply this formalism to calculate the peak profile in the Gaussian and non-Gaussian limits. In Sec. \ref{sec:results} we present our results and in Sec. \ref{sec:conclusion} we draw the final conclusions.

\section{Semianalytic model for the dark matter halo profile}
\label{sec:halo_model}
Dark matter halos form from the peaks of the initial density field. In the so-called peak formalism, one assumes that gravitationally bound objects form at local maxima (peaks) of the linear density field. Therefore the material that collapsed to form a halo of a given mass can be identified in the initial linear density field first by smoothing it with a filter of appropriate scale and then locating all the peaks above some threshold. Using the main assumption of the peak formalism, Dalal {\it{et al.}} \cite{Dalal:2010hy} suggested that the density profile within a halo can be determined by applying the spherical collapse model to the spherically averaged profile of the peak of the linear density field that collapses to form that halo.

The statistics of the peaks and the profile of the density field in their vicinity were derived for Gaussian fields in a seminal paper by Bardeen, Bond, Kaiser and Szalay (BBKS) \cite{Bardeen:1985tr}. It is convenient to define the overdensity\footnote{We implicitly center the peak at $x = 0$, so $\delta_{\text{\rm pk}} \equiv \bar\delta({\bf x};r_{\text{\rm pk}}) = \bar\delta ({\bf x} = 0;r_{\text{\rm pk}})$.} smoothed on a scale $r$ as
\begin{eqnarray}
\bar\delta({\bf x};r) &=& \int d^3 x' \ W(| {\bf x}-  {\bf x}'|; r) \delta({\bf x}')  \cr
&=& \int \frac{d^3k}{(2\pi)^3} \tilde W(k;r) \delta({\bf k}) e^{i{\bf k}.{\bf x}},
\end{eqnarray}
where the smoothing function is a top hat in real space so that 
\begin{equation}
\tilde W(k;r) = \frac{3(\text{sin}(kr) - kr \text{cos}(kr))}{(kr)^3}.
\end{equation} 
For a spherically averaged peak on a scale $r_{\text{\rm pk}}$ that collapses to form a halo of mass $M\simeq (4\pi/3)\bar \rho r_{\text{\rm pk}}^3$, where $\bar\rho$ is the mean density, the statistics of the linear density field in the inner regions on scales $r_L < r_{\text{\rm pk}}$ can be calculated given the height of the peak, $\bar \delta (r_{\text{\rm pk}}) \equiv \delta_{\text{\rm pk}}$ and the derivative of the linear density field on this scale, $d \bar \delta/dr_L  |_ {\text{\rm pk}}  \equiv \delta '_{\text{\rm pk}}$. The key quantity then is the conditional probability $P( \bar \delta(r_L)|\delta_{\text{\rm pk}},\delta'_{\text{\rm pk}})$.  For Gaussian variables, $X$ and $Y$, the conditional probability $P(X|Y)$ is also a Gaussian. Taking
\begin{align}
X \equiv \bar \delta(r_L), \qquad \qquad  \qquad
Y \equiv \begin{pmatrix}
    \delta_{\text{\rm pk}} \\ \\
    \delta'_{\text{\rm pk}}
\end{pmatrix},
\end{align}
the mean of $\bar\delta$ given these boundary conditions is
\be
\hspace{-1in} \langle X|Y\rangle \ = \ \langle XX\rangle\langle YY\rangle^{-1} Y,
\eql{mean}
\ee
where $\langle YY\rangle$ is a $2\times2$ matrix with, e.g., $\langle YY\rangle_{12} = \langle \delta_{\text{\rm pk}} \delta'_{\text{\rm pk}}\rangle$.
The variance of the fluctuations is
\be
\sigma_{X|Y} ^2 \ = \ \langle XX\rangle- \langle XY\rangle \langle YY\rangle ^ {-1} \langle Y X \rangle . 
\eql{variance}\end{equation}

Therefore, in the Gaussian case, the power spectrum for the initial linear density field $\delta({\bf k})$ directly determines the mean and variance of $\bar\delta(r_L)$, the overdensity in the interior of the peak. We will show in Sec. \ref{sec:con_prob} that the conditional probability $P( \bar \delta(r_L)|\delta_{\text{\rm pk}},\delta'_{\text{\rm pk}})$ can also be calculated using the path-integral formulation of excursion set theory introduced in Refs. \cite{Maggiore:2009rv,Maggiore:2009rw, Maggiore:2009rx}. This formalism is particularly useful in calculating the above conditional probability for non-Gaussian fluctuations which in turn would allow us to study the impact of primordial non-Gaussianity on the density profile of dark matter halos.

\newcommand\rp{r_{\text{\rm pk}}}

Dalal {\it{et al.}}~\cite{Dalal:2010hy} pointed out that recovering the internal profile requires another step.
The hierarchy of peaks within peaks expected for CDM cosmologies modifies the peak profile calculated from the above probability distribution. The material in the center of the final halo typically originates not from the central region of the corresponding peak but instead from a subpeak within the main progenitor. The mass within this subpeak is dragged to the center of the halo by processes like dynamical friction. They suggest that this effect can be taken into account by simply grafting the density of the highest subpeak onto the overall peak profile. Therefore the interior density profile of the peak at a given scale $r_L$ is given by the largest value of $\bar \delta$ for all the subvolumes of size  $r_{\text{sub}} = r_L$. The initial peak on scale $\rp$ contains $N$ regions of size $r_L$, where
$N = \left(r_{\text{\rm pk}}/r_L\right)^3$. The density in each of these subregions is also drawn from the same Gaussian distribution, so the probability that the density in any one subpeak is less than a given value $\bar \delta(r_L)$ is given by the cumulative distribution function
\begin{equation}
P_1(\bar \delta(r_L)) = \int _{-\infty}^{\bar \delta(r_L)} P(\delta(r)|\delta_{\text{\rm pk}},\delta'_{\text{\rm pk}}) d \delta(r).
\eql{p1}
\end{equation}
Therefore the probability that the density of all $N$ subpeaks is less than this value is given by 
\begin{equation}
\label{eq:peak_prof}
P_N(\bar \delta(r_L)) = P_1(\bar \delta(r_L))^N.
\end{equation} 
Therefore the probability that a subpeak at a scale $r_L$ has the highest density among all the subpeaks is given by $d P_N/d \ \bar\delta(r_L)$. From this probability distribution, we can compute the mean and variance of the full set of subhalos. 

Given the profile of the initial peak in Eq. (\ref{eq:peak_prof}), Ref.~\cite{Dalal:2010hy} took a further step to relate this profile to the mass and density profile of the final halo. They suggested that the final halo profile can be predicted by applying spherical collapse to the above spherically averaged profile of the initial peak. They argued that the shell crossing can be accounted for by modeling the mass profile deposited by a given mass shell within the initial peak. In the rest of this section we  briefly review their reasoning. 

Considering the profile of the peak of linear density, the overdensity on a given scale $r_L$ grows until it reaches the turnaround point and then it collapses. The turnaround point can be calculated by applying the spherical collapse model to the mass shell on this scale. Subsequent to turnaround, the mass within a given mass shell is no longer constant since the particles within it can cross the shell. Therefore the mass shell on a given scale $r_L$ with mass $dM_L$ does not deposit material over a thin shell in the final halo but instead lays down material over a range of radii, $M_s(r) = dM_L f(r)$, where $M_s(r)$ is the mass profile in the final halo laid down by a shell of width $dM_L$. This deposited mass not only increases the mass within the shell with the smaller radii but also induces a contraction of the material already present in that shell.  For a spherically symmetric object, the radial action $J(r) \propto [r \times M(r)]^{1/2}$ is conserved. Therefore if the mass at radius $r$ is increased, the radius will be decreased, an effect referred to as adiabatic contraction. This contraction can be parametrized by modeling the mass profile deposited by each mass shell of the initial density field. Two toy models for the profile of the deposited mass were introduced in Ref. ~\cite{Dalal:2010hy}, referred to as minimal and nonminimal models hereafter. The minimal contraction corresponds to the case that the shell profiles behaves as $\rho \rightarrow {\rm const}$ as $r \rightarrow 0$, while in the nonminimal model, the shell profiles have inner slopes $d\left({\rm log} \ \rho_{\rm shell}\right)/d\left({\rm log} \  r \right) \sim \frac{1}{2}d \left({\rm log} \  \rho_{\rm tot} \right)/d\left({\rm log} \ r\right)$.  The minimal model provides a lower limit on the effect of adiabatic contraction. If the shell profiles have more mass than is assumed in this model, the contraction can be stronger. The nonminimal model is an example of a model in which the shells have nonminimal tails.  As noted in Ref. \cite{Dalal:2010hy} neither of these models should be taken as a precise description of the shell profiles; rather, they are illustrative since they simplify the calculations.

Since the contraction keeps the radial action invariant, the mass profile deposited in the final halo by a given mass shell can be calculated from the value of $r \times M(r)$  before the collapse, for instance, at turnaround
\begin{eqnarray}
\label{eq:F}
 F(M_L) &\equiv& M_L \times r_{ta}(M_L)
 \vs
&=&0.6 \left(\frac{3}{4 \pi\bar\rho}\right)^{1/3} \frac{M_L^{4/3}}{\bar \delta_{\text{lin}}(M_L) },
\end{eqnarray}
where we have used $r_{\rm{ta}} \simeq  0.6 \ r_L/{\bar \delta_{\rm{lin}}}$. The mass profile of the final halo can then be obtained by integrating over the mass profiles deposited by the mass shells on scales $0<r_L<r_{\rm pk}$. For the minimal model, they show that the mass profile can be calculated by solving the ordinary differential equation
\begin{equation}
\label{eq:minimal}
 \frac{d M }{d r} = \frac{3}{r}[M - F^{-1}(Mr)],
\end{equation}
while for the nonminimal model, following the same steps, it is straightforward to show that the mass profile of the halo satisfies the following equation 
\begin{equation}
\label{eq:nonminimal}
 \frac{d M}{d r} = \frac{3 M}{r} \frac{M-F^{-1}(Mr)}{M+F^{-1}(Mr)}.
\end{equation}
In the above equations $F^{-1}$ is the inverse of the function $F$ given in Eq. (\ref{eq:F}).

\section{Path-integral formulation of excursion set theory}
\label{sec:path_integral}
As described in Sec. \ref{sec:halo_model}, the conditional probability $P( \bar \delta(r_L)|\delta_{\text{\rm pk}},\delta'_{\text{\rm pk}})$ is necessary to compute the final density profile of dark matter halos. In the Gaussian case, the conditional probability can be equivalently computed using the path-integral formulation of excursion set theory and one can reproduces Dalal {\it et al} 's result. This formalism is particularly useful in extending the Gaussian results to the non-Gaussian case. Therefore in this section we briefly review the path-integral formulation of excursion set theory and in Sec. \ref{sec:con_prob} we calculate the conditional probability in the Gaussian and non-Gaussian limits. 

Excursion set theory was introduced by Bond, Cole, Efstathiou and Kaiser \cite{Bond:1990iw} (see Zentner \cite{Zentner:2006vw} for a recent review). It provides an alternative derivation of the Press-Shechter mass function while solving the so called ``cloud-in-cloud'' problem of the original derivation of Press and Shechter. The path-integral formulation of excursion set theory was developed by Maggiore and Riotto (MR) in a series of papers~\cite{Maggiore:2009rv,Maggiore:2009rw, Maggiore:2009rx}. This formalism allows for straightforward extensions of  excursion set theory to the cases of non-Gaussian initial conditions, moving barrier, and top-hat filter in real space. 

Following MR, we consider an ensemble of trajectories of $\delta(S)$ all starting from the same initial point $\delta(S_0 )$ and follow them for a time $S$. This  variable $S$ is in one-to-one correspondence with the smoothing scale $r$ by the relation
$S=\langle\bar{\delta}^2({\bf x},r)\rangle$.
We discretize the time interval $[0,S]$ into infinitesimal steps, $\Delta S = \epsilon$, so that $S_k = k \epsilon$ where $k = 0,1,..,n $ and $S_n \equiv S$. The discretized trajectory is then defined as a set of values ${\delta_1,\delta_2,..,\delta_n}$ where $\delta(S_i) \equiv\delta_i$. The probability of arriving at point $\delta_n$ at time $S$ starting from the initial point $\delta_0$ through trajectories that never exceeded the threshold is given by
\begin{equation}
\Pi(\delta_0;\delta_n;S_n) \equiv \int_{-\infty}^{\delta_c} d\delta_1 \hdots  \int_{-\infty}^{\delta_c} d\delta_{n-1} W(\delta_0;...;\delta_n;S_n),
\end{equation}
where 
\begin{align}
W(\delta_0;...,\delta_n;S_n) &\equiv \langle \delta_D(\delta(S_0)-\delta_0) ... \delta_D(\delta(S_n)-\delta_n)\rangle \nonumber \\[5pt]
& 
\end{align}
is the probability density in the space of trajectories. To avoid confusion with the overdensity $\delta$, the Dirac delta function is denoted as $\delta_D$. 

A more useful expression for the probability distribution $\Pi(\delta_0;\delta_n;S_n)$ can be derived by writing the probability density function $W(\delta_0;..;\delta_n;S_n)$ in terms of $p$-point correlation functions of $\delta$. The trick is to use the integral representation of Dirac delta function,
\begin{equation}
\delta_D(x) = \int _{-\infty}^\infty \frac{d\lambda}{2\pi}e^{-i\lambda x},
\end{equation}
so that 
\begin{equation}
\label{eq:prob_dens}
W(\delta_0;...;\delta_n,S_n) = \int_{-\infty}^\infty {\mathcal D} \lambda \ e^{\sum_{i=0}^n \lambda_i\delta_i} \langle e^{-i \sum_{i=0}^n \lambda_i \delta(S_i)}\rangle, 
\end{equation}
where we defined 
\begin{equation}
\int_{-\infty}^\infty \mathcal D \lambda\equiv \int_{-\infty}^\infty\frac{d\lambda_0}{2\pi} \hdots \int_{-\infty}^\infty\frac{d\lambda_n}{2\pi}.
\end{equation}
The expectation value in Eq. (\ref{eq:prob_dens}) can then be written as
\begin{align}
\langle & e^{-i \sum_{i=0}^n  \lambda_i \delta(S_i)}\rangle  \nonumber \\
& \hspace{0.1in}= \text{exp}\left[\sum_{p=2}^\infty \frac{(-i)^p}{p!}\sum_{j_1,...,j_p=0}^n{\lambda_j}_1 \hdots {\lambda_j}_p \langle \delta({S_j}_1)...\delta({S_j}_p)\rangle_c \right],
\end{align}
where  $\langle \delta({S_j}_1)...\delta({S_j}_p)\rangle_c$ is the connected $p$-point function. For Gaussian fluctuations, all the $p$-point functions with $p>2$ vanish and the probability density reduces to
\begin{align}
\label{eq:prob_dens_G}
W_G&(\delta_0,...,\delta_n,S_n) =  \nonumber \\
 & \int_{-\infty}^\infty {\mathcal D} \lambda \ \text{exp}\left({-i\sum_{i=0}^n \lambda_i\delta_i - \frac{1}{2}\sum_{i,j=0}^n \lambda_i\lambda_j \langle \delta_i\delta_j\rangle_c} \right).
\end{align}
The non-Gaussian effects primarily arise from the nonzero contribution of the three-point correlation function. So by dropping the higher-order correlators, the probability density $W_{\rm{NG}}(\delta_0;\delta_n;S_n)$ can be written as
\begin{align}
\label{eq:prob_dens_NG}
W_{NG}&(\delta_0;\hdots;\delta_n;\delta_n) = \nonumber \\
&\int \mathcal D \lambda  \ \text{exp}  \left( i \sum_{i=0}^n \lambda_i \delta_i - \frac{1}{2} \sum_{i,j = 0}^n \langle \delta_i\delta_j\rangle_c \lambda_i \lambda_j  \nonumber \right.\\ 
& \hspace{0.5in} \left .+ \frac{(-i)^3}{6} \sum _{i,j,k=0}^n \langle \delta_i\delta_j\delta_k\rangle_c \lambda_i\lambda_j\lambda_k \right).
\end{align}
For (the realistic case of) small non-Gaussianities, considering only the three-point function, the non-Gaussian probability density can be written perturbatively as
\begin{equation}
W_{NG} \approx W_{G} - \frac{1}{6} \sum_{i,j,k=1}^n \langle \delta_i\delta_j\delta_k\rangle_c\partial_i\partial_j\partial_k W_{G}.
\eql{ng}
\end{equation} 
In order to derive the mass profile of the collapse halo, as explained in Sec. \ref{sec:halo_model}, we need to calculate the conditional probability $P(\delta_n|\delta_0,\delta_1)$. In the next section we compute this conditional probability in terms of the probability distribution $\Pi(\delta_0;\delta_n;S_n)$ of the path-integral formalism. 

\section{Conditional probability from path-integral formulation}
\label{sec:con_prob}
\noindent

Our goal is to calculate the conditional probability $P(\bar \delta(r_L)|\delta_{\text{\rm pk}},{\delta'}_{\text{\rm pk}})$, the probability of being in an overdense region $\delta_n = \bar \delta(r_L)$ in the interior region of a peak given the height of the peak $\delta_0 = \delta_{\text{\rm pk}}$ and the slope of the linear density profile at the location of the peak $\delta_{\text{\rm pk}}'$. We can also think of this as fixing the first two points in the trajectory (since the derivative is related to the difference of the first two points). The conditional probability then corresponds to the probability of arriving at a point $\delta_n$ at time $S_n$ given the first two steps,  $P(\delta_n|\delta_0,\delta_1)$,
\begin{equation}
\label{eq:con_prob}
P(\delta_n|\delta_0,\delta_1) = \frac{\Pi(\delta_0;\delta_1;\delta_n)}{\Pi(\delta_0;\delta_1)}.
\end{equation}
As a first-order approximation, we calculate the above conditional probability by considering all the trajectories between $\delta_0$ and $\delta_n$ that pass through the intermediate step $\delta_1$, including those that have passed the threshold. Therefore the integral limits vary in the range $[-\infty,\infty]$. Making this assumption is equivalent to assuming that the evolution of $\delta$ is Markovian. We will use this Markovianity when calculating the non-Gaussian conditional probability. The probability density $\Pi(\delta_0;\delta_1;\delta_n)$,  
\begin{equation}
\label{eq:prob_fixdelta1}
\Pi(\delta_0;\delta_1;\delta_n) \equiv \int_{-\infty}^{\infty} d\delta_2  \hdots  \int_{-\infty}^{\infty} d\delta_{n-1}  W(\delta_0;\delta_1;...;\delta_n),
\end{equation}
where $W(\delta_0;\delta_1;...;\delta_n)$, is given by Eq. (\ref{eq:ng}). Note that since we are fixing the intermediate step $\delta_1$, the probability density of Eq. (\ref{eq:prob_fixdelta1}) is defined in terms of a summation over intermediate steps $\delta_i$, where $i \geqslant 2$.

\subsection{Gaussian limit}
For Gaussian fluctuations, the probability density is given by Eq. (\ref{eq:prob_dens_G}). Using the integral representation of the Dirac delta function simplifies the equations since an integration over intermediate steps $\delta_i \in [\delta_2,\delta_{n-1}]$ gives unity and we have
\begin{align}
\label{eq:multivar}
 \Pi_G(\delta_0;\delta_1;\delta_n) &=  \int_{-\infty}^{\infty} \frac{d\lambda_0}{2\pi} \int_{-\infty}^{\infty} \frac{d\lambda_1}{2\pi} \int_{-\infty}^{\infty} \frac{d\lambda_n}{2\pi} \nonumber \\
 & \times \text{exp} \left\{i\sum_{i=0,1,n}\lambda_i\delta_i - \frac{1}{2} \sum_{i,j=0,1,n}\lambda_i\lambda_j \xi_{ij}\right\} , 
\end{align}
where we have defined the two-point functions as $\xi_{ij}\equiv\langle\delta_i\delta_j\rangle$, so, for example, $S_0=\xi_{00}$. Carrying out the Gaussian integrals leads to 
\begin{equation}
\Pi_G(\delta_0;\delta_1;\delta_n)= \sqrt{\frac{(2\pi)^3}{\det(\Xi_1)}}  \text{exp}\left(-\,\frac{1}{2}  \vec\Delta_1^T   {\Xi_1}^{-1}  \vec{\Delta_1}\right),
\end{equation}
where we have defined 
\begin{align}
\Xi_1 &\equiv \begin{pmatrix}
        S_0  & \xi_{01} & \xi_{0 n}  \\ 
        \xi_{01}  &  S_1 &   \xi_{1 n}  \\ 
        \xi_{0 n}  &  \xi_{1 n} &  S_n
\end{pmatrix},
&
\vec \Delta_1 &\equiv \begin{pmatrix}
                \delta_0  \\
                 \delta_1  \\
                 \delta_n
\end{pmatrix}.
\end{align}
Similarly, $\Pi_G(\delta_0;\delta_1)$ is given by
\begin{equation}
\Pi_G(\delta_0;\delta_1)= \sqrt{\frac{(2\pi)^3}{\text{det} \ \Xi_2}} \ \text{exp}\left(-\,\frac{1}{2} \ \vec {\Delta_2}^T  \ {\Xi_2}^{-1} \ \vec \Delta_2\right),
\end{equation}
with
\begin{align}
\Xi_2 &= \begin{pmatrix}
        S_0 & \xi_{0 1}  \\ 
        \xi_{01}  & S_1  \\ 
\end{pmatrix},
&
\vec \Delta_2 &= \begin{pmatrix}
                \delta_0  \\
                 \delta_1  \\
\end{pmatrix}.
\end{align}
Therefore the conditional probability in the Gaussian limit is 

\begin{align}\label{eq:con_prob_G_ex}
\hspace{-0.1in} P_G(\delta_n|\delta_0;\delta_1) &= \frac{\Pi_G(\delta_0;\delta_1;\delta_n)}{\Pi_G(\delta_0;\delta_1)} \nonumber \\[2pt]
& = \sqrt{\frac{\text{det} \ \Xi_2}{\text{det} \ \Xi_1}} \ e^{-\,\left[\frac{1}{2} \left( \vec {\Delta_1}^T   {\Xi_1}^{-1}  \vec \Delta_1  - \vec {\Delta_2}^T   {\Xi_2}^{-1}  \vec \Delta_2\right)\right]}.
\end{align}

This equation was used to generate the results shown in Fig.~\rf{peak_G}, which agree with those in Ref.~\cite{Dalal:2010hy}. Indeed, it is clear from Eq. \ref{eq:con_prob_G_ex} that the conditional probability for $\delta_n$ is Gaussian, and a little algebra shows that the mean and variance agree with Eqs.~\Ec{mean} and \Ec{variance}.

\subsection{Non-Gaussian limit}
For non-Gaussian fluctuations the probability density is given by \ec{ng}, so
\bea
\Pi_{NG}(\delta_0; \delta_1;\hdots;\delta_n)  &\simeq&  \int d\delta_2\,\int d\delta_3\ldots \int d\delta_{n-1} \vs
\times  \Big(1 -\frac{1}{6}\sum_{i,j,k=0}^n &\langle\delta_i\delta_j\delta_k\rangle_c &\partial_i\partial_j\partial_j \Big)
W_G(\delta_0;\delta_1;\hdots;\delta_n).
\vs
\eea
with a similar expression for $\Pi(\delta_0;\delta_1)$ except $\delta_n$ is also integrated over. 
The conditional probability is then 
\begin{equation}
P_{NG}(\delta_n|\delta_0,\delta_1) = \frac{\Pi_{\rm{NG}}(\delta_0;\delta_1;\delta_n)}{\Pi_{\rm{NG}}(\delta_0;\delta_1)}.
\end{equation}

As discussed before, to calculate the conditional probability, we make the assumption that for Gaussian fluctuations the evolution of $\delta$ as a function of $S$  is Markovian. This allows us to write
\begin{equation}
\label{eq:Markovian}
W_G(\delta_0;\delta_1;\hdots;\delta_n) = W_G(\delta_0;\delta_1)W_G(\delta_1;\hdots;\delta_n).
\end{equation}
In  $\Pi_{\rm{NG}}(\delta_0;\delta_1;\hdots;\delta_n;S_n)$ the non-Gaussian contributions  can be divided into those that depend on the end steps ($\delta_0,\delta_1,\delta_n$) and those that do not,
\begin{align}
\label{eq:NG_terms}
\sum_{i,j,k=0}^n \langle \delta_i\delta_j\delta_k\rangle \ \partial_i\partial_j\partial_k  \  &= \sum _{i,j,k=0}^1 \langle \delta_i\delta_j\delta_k  \rangle_c \ \partial_i\partial_j\partial_k \nonumber \\ 
& + \  3 \ \sum _{i,j=0}^1  \  \langle \delta_i\delta_j\delta_n\rangle_c \  \partial_i\partial_j\partial_n \nonumber \\ 
& + \ 3 \ \sum _{i=0}^1 \  \langle \delta_i {\delta_n} ^2 \rangle_c  \ \partial_i{\partial_n}^2 \nonumber \\
&+ \ \langle {\delta_n}^3 \rangle_c \  {\partial_n}^3 \nonumber \\
&+ \  \hdots ,
\end{align}
where the dots stand for all the terms that involve at least one index $2 \leqslant i \leqslant n-1$. These terms vanish upon integration and therefore do not contribute to
$\Pi(\delta_0;\delta_1;\hdots;\delta_n)$. Moreover, using the Markovian property of Eq. (\ref{eq:Markovian}), we see that the contribution from the first term in Eq. (\ref{eq:NG_terms}) is cancelled by the equal term in 
$\Pi(\delta_0;\delta_1)$. Therefore this term does not contribute to the conditional probability 
$P(\delta_n|\delta_0;\delta_n)$. The conditional probability therefore reduces to
\begin{equation}
\label{eq:con_prob_NG}
P_{NG}(\delta_n|\delta_0\delta_1) = P_G(\delta_n|\delta_0,\delta_1) + \Delta P_{NG}(\delta_n|\delta_0,\delta-1),
\end{equation}
where
\begin{align}
\label{eq:NG-corrections}
\Delta P_{NG} (\delta_n|\delta_0,\delta_1) = &-\frac{1}{2} \sum_{i,j=0}^1 \langle \delta_i\delta_j\delta_n  \rangle_c \partial_i \partial_j\partial_n P_G(\delta_n|\delta_0,\delta_1) \nonumber \\
&-\frac{1}{2}\sum_{i=0}^1\langle \delta_i {\delta_n}^2\rangle_c \partial_i{\partial_n}^2 P_G(\delta_n|\delta_0,\delta_1) \nonumber \\
& - \frac{1}{6} \langle {\delta_n}^3 \rangle_c {\partial_n}^3 P_G(\delta_n|\delta_0,\delta_1).
\end{align}
In Sec. \ref{sec:results}, we evaluate the non-Gaussian corrections numerically, but first we present an analytic approximation.

\subsection{Large-$\delta$ approximation}

In the limit that $\delta$ is large, a limit that holds as one moves to smaller smoothing scales, we can obtain a simple analytic estimate for the most likely value of $\delta$ and for the non-Gaussian correction. Recall that $P_N(\delta)$ gives the cumulative probability that the overdensity of all subhalos is smaller than $\delta$. The derivative $dP_N/d\delta$ is the differential probability that the largest overdensity is equal to $\delta$. The most likely value then is the value of $\delta$ at which $dP_N/d\delta$ peaks.  Equivalently, the most likely value of $\delta$ is fixed by setting
\be
\frac{d^2}{d\delta^2} P_1^N = 0\eql{detbar}
.\ee
Carrying out the derivatives and setting $N-1\rightarrow N$ leads to
\be
N \left(P_1'\right)^2 + P_1 P_1'' =
0.\eql{approx}
\ee
But, by definition \ec{p1}, $P_1'=P$ and, since we are interested in the large-$\delta$ limit, $P_1$ is extremely close to one, 
so \ec{approx} reduces to
\be
N P^2 = -P'
\eql{start}.\ee
We can first solve this in the Gaussian limit. For large $(\delta_n,S_n)$, Eq. (\ref{eq:con_prob_G_ex}) reduces to
\be
P_G \rightarrow \frac{1}{\sqrt{S_n}} \exp \left\{ -\delta^2/2S_n \right\}
.\ee
Therefore, the mean value of $\delta_n$ on small scales is given by the equation
\be
N e^{-\bar\delta_n^2/2S_n} = \frac{\bar\delta_n}{\sqrt{S_n}}
.\ee
This is what we expect: the exponential suppression is offset by the large number of sub-regions that can attain independent values. Since $N=(r_{\rm pk}/r_L)^3$, to a good approximation
\be
\bar\delta_n \simeq \sqrt{2S_n} \left(\ln (r_L/r_{\rm pk})^3 \right)^{1/2}
.\ee
The mean value of $\delta$ on small scales, in this simple approximation, does not depend on the boundary conditions (on the values of the overdensity smoothed on large scales). Physically, this is reasonable: the fluctuations in the perturbations on very small scales are so large that they are virtually independent of the large-scale density field. The resulting numerical value for $\delta_n$ is within 10\% of the value plotted in Fig.~1 when $r_L=10^{-3}r_{\rm pk}$.

Now we consider the non-Gaussian correction; \ec{start} generalizes to
\be
N (P+\Delta P_{NG})^2 = -(P+\Delta P_{NG})'
\ee
By expanding this about the zero-order solution, $\delta \rightarrow \delta+ \delta^{\rm{NG}}$, and keeping
terms linear in $\delta^{\rm{NG}}$ and $\Delta P_{\rm{NG}}$, we obtain
\be
2NP\left[ P'\delta^{NG} + \Delta P_{NG}\right] = - P'' \delta^{NG} - \Delta P_{NG}'
\ee
So this immediately determines the shift due to non-Gaussianity,
\be
\delta^{\rm{NG}} = - \frac{\Delta P_{\rm{NG}}' + 2NP_G\Delta P_{\rm{NG}} }{P_G''+2NP_GP_G'}\eql{fo}
\ee
The denominator can be rewritten using our zero-order solution as
\be
P_G''+2NP_GP_G' = -\frac{P}{S_n}\left( 1 + \frac{\delta^2}{S_n} \right),
\ee
and the numerator reduces to
\be
\Delta P_{NG}' + 2NP_G\Delta P_{NG}=
\frac{\delta^2}{6S_n^3} \langle {\delta_n}^3 \rangle_c P_G
\left[ 3 + \frac{\delta^2}{S_n} 
\right]
.\ee
This leads to a simple expression for the change in the mean value of $\delta$ due to non-Gaussianity,
\be
\label{eq:largedelta}
\delta^{NG}=
\frac{\delta^2\langle {\delta_n}^3 \rangle_c}{6S_n^2} \frac{
\left[ 3 + \frac{\delta^2}{S_n} 
\right]}{\left( 1 + \frac{\delta^2}{S_n} \right)}
\rightarrow
\frac{\delta^2\langle {\delta_n}^3 \rangle_c}{6S_n^2} .
\ee
We will see in the next section that this approximation is not quite as good as the Gaussian approximation, overshooting by about 40\%.
\begin{figure}
\vspace{-20pt}
\includegraphics[width= 3.3in]{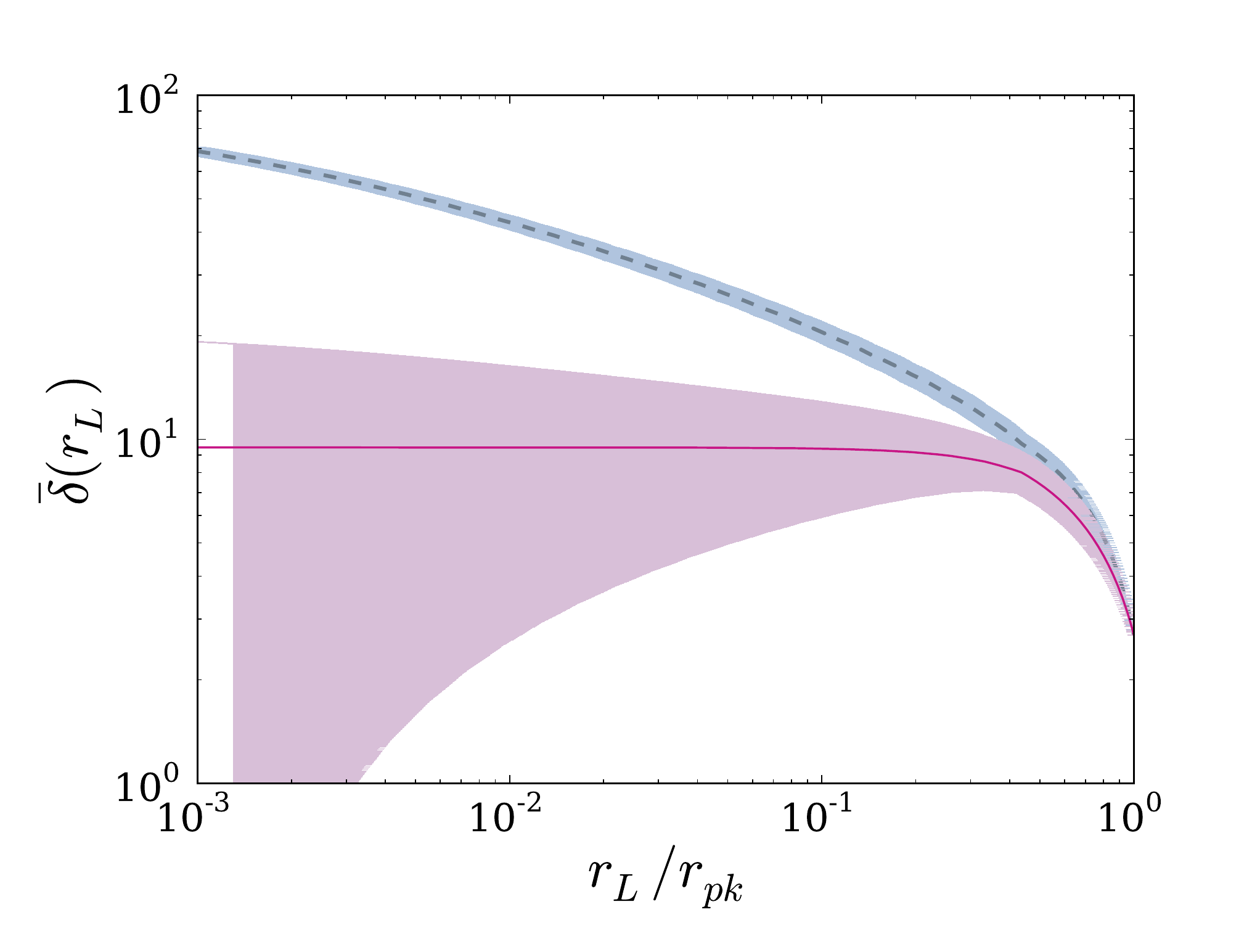}
\vspace{-10pt}\caption{Initial linear density profile of the peak. $r_{\rm pk}$ is the scale of the initial peak which collapses to form the halo while $r_L$ corresponds to the radius of the initial Lagrangian shell. The purple solid line is the mean of the density field when conditioned on the first two steps: the height of the central peak $\delta_{\rm pk}$ and an intermediate step $\delta_1$. The gray dashed line is the mean of the density field when the off-center peaks are taken into account. The shaded regions are the corresponding dispersions.}
\label{fig:peak_G}
\end{figure}
\vspace{-10pt}

\section{Results}
\label{sec:results}
\begin{figure*}[t]
\includegraphics[width=\columnwidth]{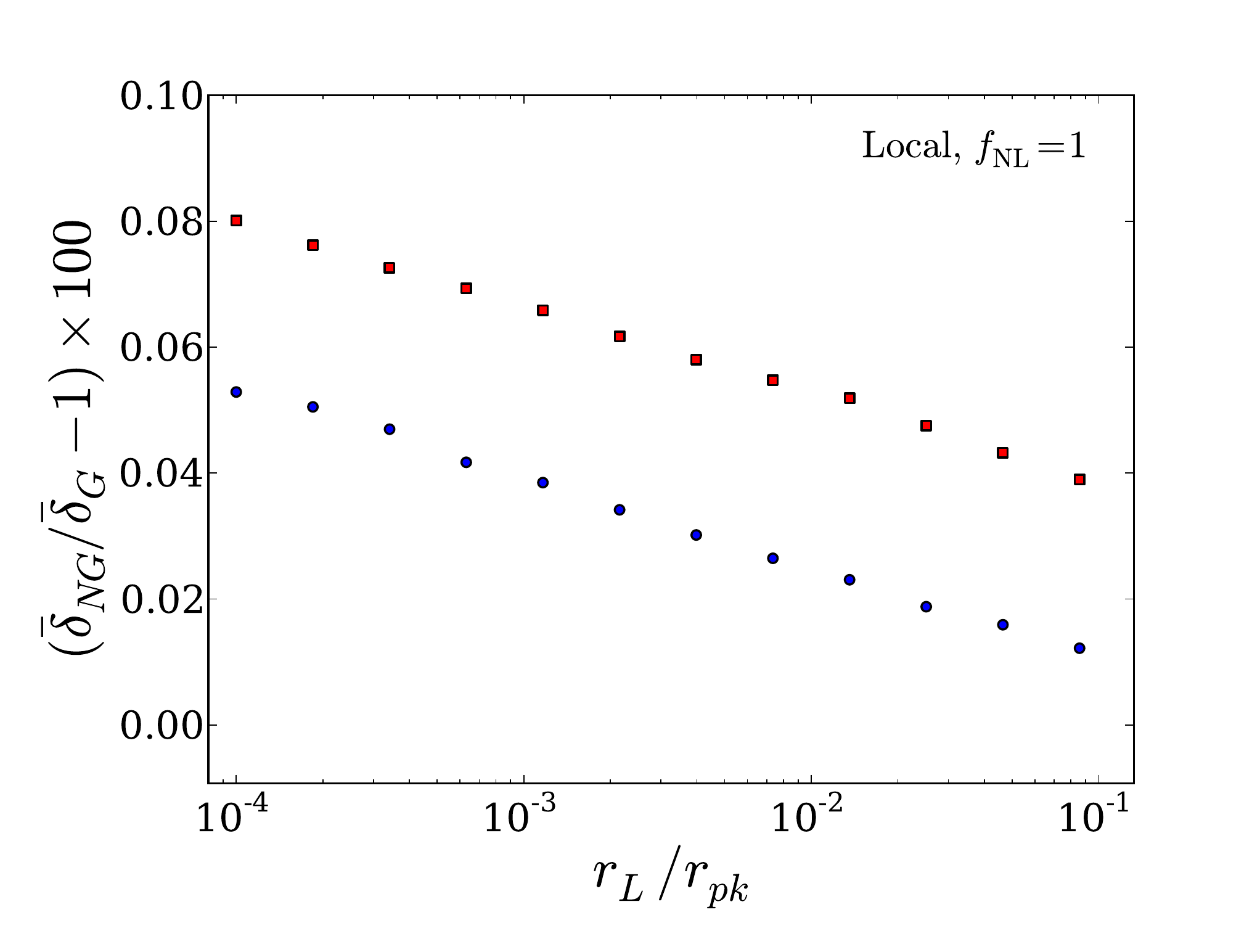}
\includegraphics[width=\columnwidth]{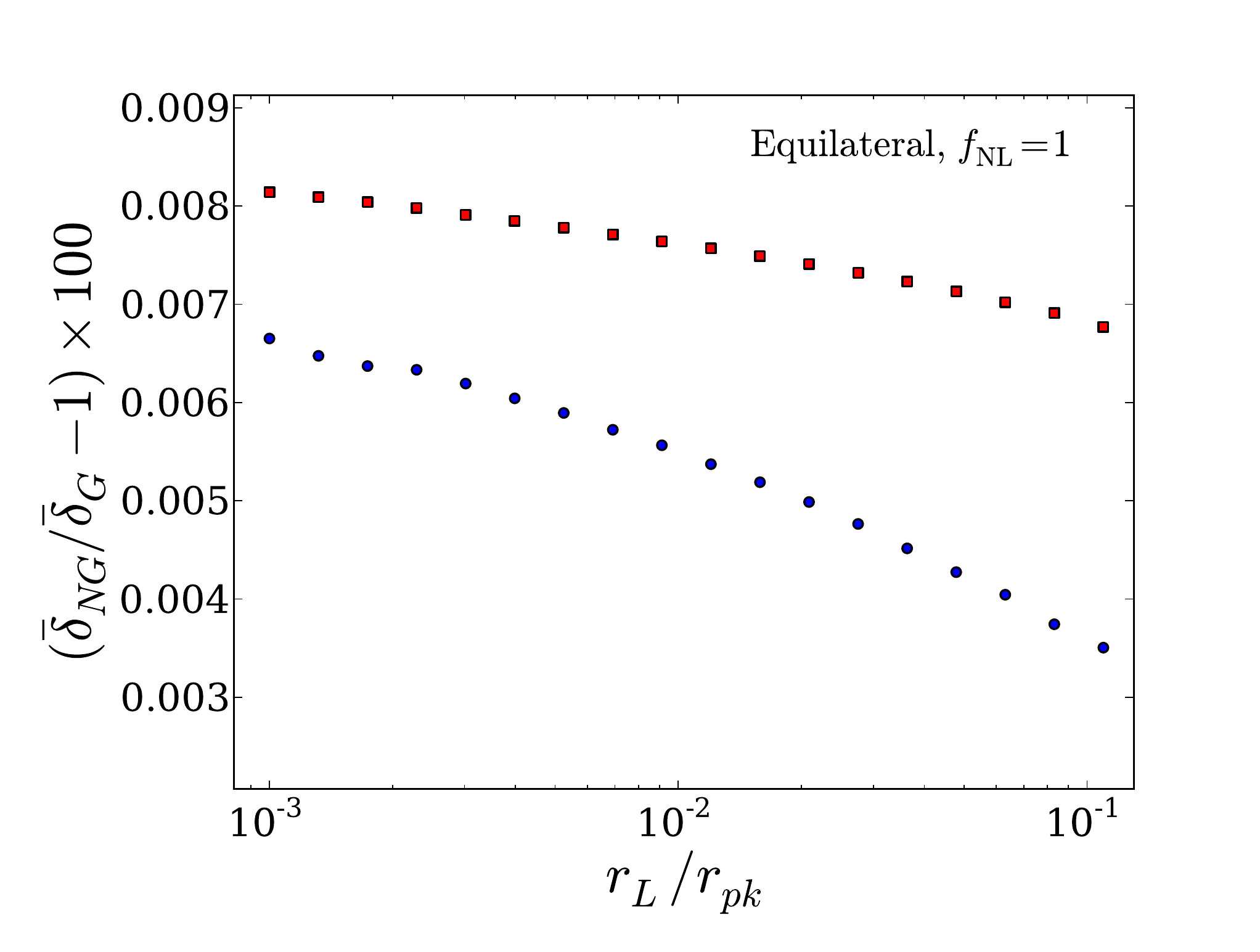}
\caption{Ratio of mean of the highest subpeak profile for non-Gaussian fluctuations with $f_{\rm NL} =1$ to that of Gaussian fluctuations. The left panel corresponds to local-shape while the right panel corresponds to equilateral-shape non-Gaussianity. In both plots, the blue circles are the numerical calculation of the non-Gaussian corrections while the red squares correspond to the large $\delta$ approximation of Eq. \ref{eq:largedelta}.}
\label{fig:mean_ratio}
\end{figure*}

In this section we first calculate the linear density profile of the initial peak (the BBKS profile) in the Gaussian limit using Eq. (\ref{eq:con_prob_G_ex}). Next we calculate non-Gaussian corrections to the peak profile by evaluating Eq. (\ref{eq:con_prob_NG}) numerically. These conditional probabilities are then used to calculate the statistics of the highest subpeaks from the differential probability $dP_N/d\delta$, where $P_N$ is given by Eq. (\ref{eq:peak_prof}). Finally, we use Dalal {\it et al.} 's prescription [Eq. (\ref{eq:minimal}) or Eq. (\ref{eq:nonminimal})] to relate the peak profile to the final density profile of the halo. 

We choose the initial conditions such that the peak profile in the Gaussian limit, agrees approximately with that in Fig. 2 of Ref. \cite{Dalal:2010hy}: 
the height of the peak is taken to be $\delta_{\rm pk} = \delta_0 = 2.7$ and the density field at the intermediate step is $\delta_1 = 3.73$. These initial conditions approximately correspond to the main halo in the high-resolution Via Lectea II simulations by Diemand {\it et al.} \cite{Diemand:2008in}. The simulation begins at redshift of $z=104.3$ and outputs 400 snapshots in time ending at $z=0$. The halo properties are determined by processing the snapshots. At $z=0$ the main halo has $r_{200} = 402 \ \text{kpc}$ and $M_{200} = 1.92 \times 10^{12} M_{\odot}$ where $r_{200}$ is the radius enclosing a density of $200 \rho_M = 200 \Omega_M\rho_{\rm{crit}}$ and $M_{200}$ is the the corresponding mass. The scale of the peak $r_{\text{\rm pk}}$ is given by $M_{200} = (4\pi/3)\bar \rho r_{\rm{pk}}^3$.

Figure \ref{fig:peak_G} shows the mean and dispersion of the linear density field in the vicinity of the peak for Gaussian fluctuations. The purple solid line is the mean while the purple shaded area is the dispersion for the BBKS profile. Including the hierarchy of peaks within peaks the mean and dispersion of the density profile around the peak is modified. The grey dashed line and the shaded grey area show the mean and dispersion in this case. This figure can be compared with Fig. 2 in Ref. \cite{Dalal:2010hy}. It should be noted that the slight offset in the value of mean in the two plots is due to small differences in the initial conditions.

To calculate the non-Gaussian corrections to the halo mass profile, we first need to calculate the non-Gaussian conditional probability given in Eq. (\ref{eq:con_prob_NG}). This requires calculating the three-point correlator. In general the three-point function in Fourier space is given by
\begin{align}
\label{eq:3-point}
\langle \delta_i \delta_j \delta_k \rangle_c = &\int \frac{d^3 {\bf k}_1}{(2\pi)^3}  \int \frac{d^3 {\bf k}_2}{(2\pi)^3} \int \frac{d^3 {\bf k}_3}{(2\pi)^3} W(k_1,r_i)W(k_2,r_j) \nonumber \\
&\times W(k_3,r_k) \langle \delta({\bf k}_1)\delta({\bf k}_2) \delta({\bf k}_3) \rangle e^{-i({\bf k}_1+{\bf k}_2+ {\bf k}_3).{\bf x}}
\end{align}
The above equation can be written in a more convenient form in terms of the gravitational potential. The present-day density field is related to the primordial value of the gravitational potential $\Phi$ via
\begin{equation}
\delta({\bf k},a_0) = \alpha(k,a_i,a_0)\Phi({\bf k},a_i),
\end{equation} 
where
\begin{equation}
\alpha(k ,a_i,a_0) \equiv \frac{2g(a_i,a_0)k^2T(k,a_0)}{3\Omega_m {H_0}^2}. 
\end{equation}
Here $g(a_1,a_2) \equiv[D(a_2)/D(a_1)][a_1/a_2]$ is the growth suppression factor ($a_1<a_2)$. For $\Lambda \text{CDM}$ cosmology $g(a_i,a_0) \approx 0.75$. $T(k,a_0)$ is the matter transfer function normalized to unity as $k \rightarrow 0$. Defining the three-point function of the gravitational potential  $\Phi$ in terms of the bispectrum, 
\begin{equation}
\langle \Phi({\bf k}_1) \Phi({\bf k}_2) \Phi({\bf k}_3) \rangle  = B_\Phi (k_1,k_2,k_3) (2\pi)^3\delta^3({\bf k}_1+{\bf k}_2+{\bf k_3}), 
\end{equation}
Eq. (\ref{eq:3-point}) reduces to
\begin{align}\label{eq:3point}
\langle\delta_i\delta_j\delta_k\rangle_c &= \int_0^\infty \frac{dk_1 {k_1}^2\alpha(k_1)}{2\pi^2} \int_0^\infty \frac{ dk_2 {k_2}^2\alpha(k_2)}{4\pi^2} \int_{-1}^1  d\mu   \nonumber \\ 
\times &\alpha(k_3)W(k_1,r_i)W(k_2,r_j)W(k_3,r_k) B_\Phi({k}_1,{k}_2,{k}_3), 
\end{align}
\begin{figure*}[t]
\includegraphics[width=\columnwidth]{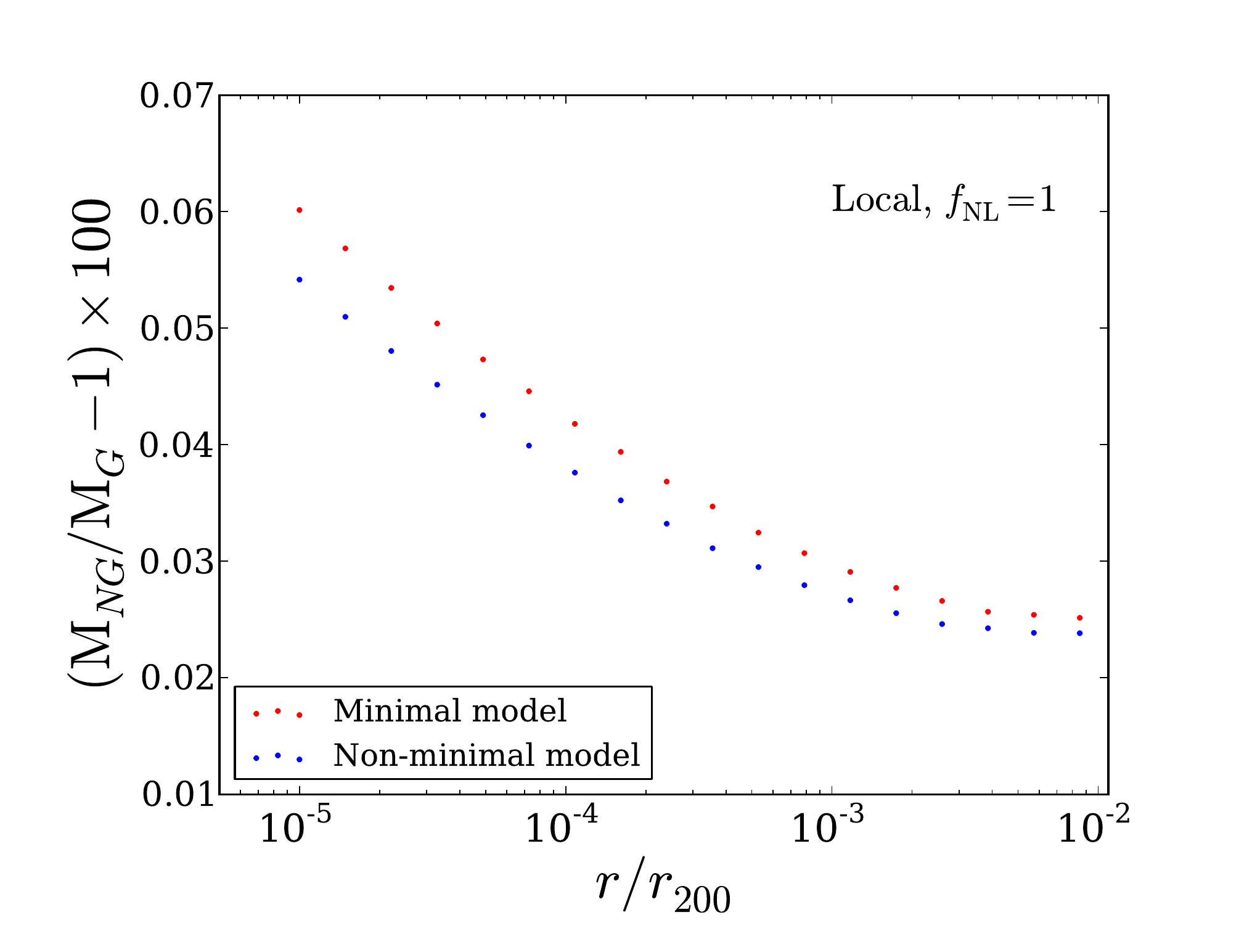}
\includegraphics[width=\columnwidth]{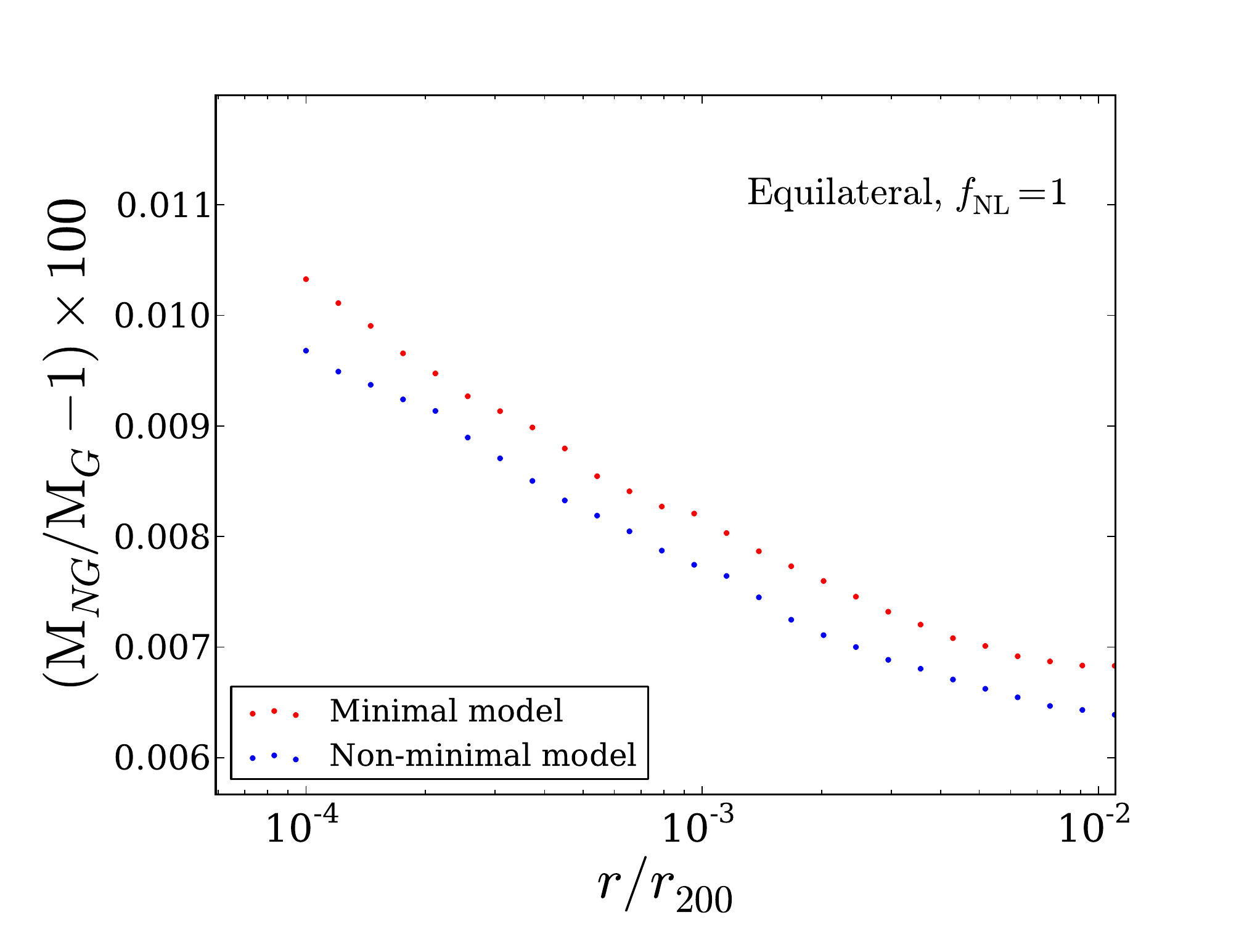}
\caption{The ratio of mass profile of the halo for non-Gaussian fluctuations 
with $f_{\rm NL}=1$ to that of the Gaussian fluctuations. The left panel corresponds to the local  shape while the right panel corresponds to the equilateral shape. In both plots, the red curve (the upper curve) shows the ratio for the minimal model of Eq. (\ref{eq:minimal}) while the blue curve (the lower curve) shows the ratio in the nonminimal model of Eq. (\ref{eq:nonminimal}).}
\label{fig:mass_ratio}
\end{figure*}
\begin{figure*}
\includegraphics[width=\columnwidth]{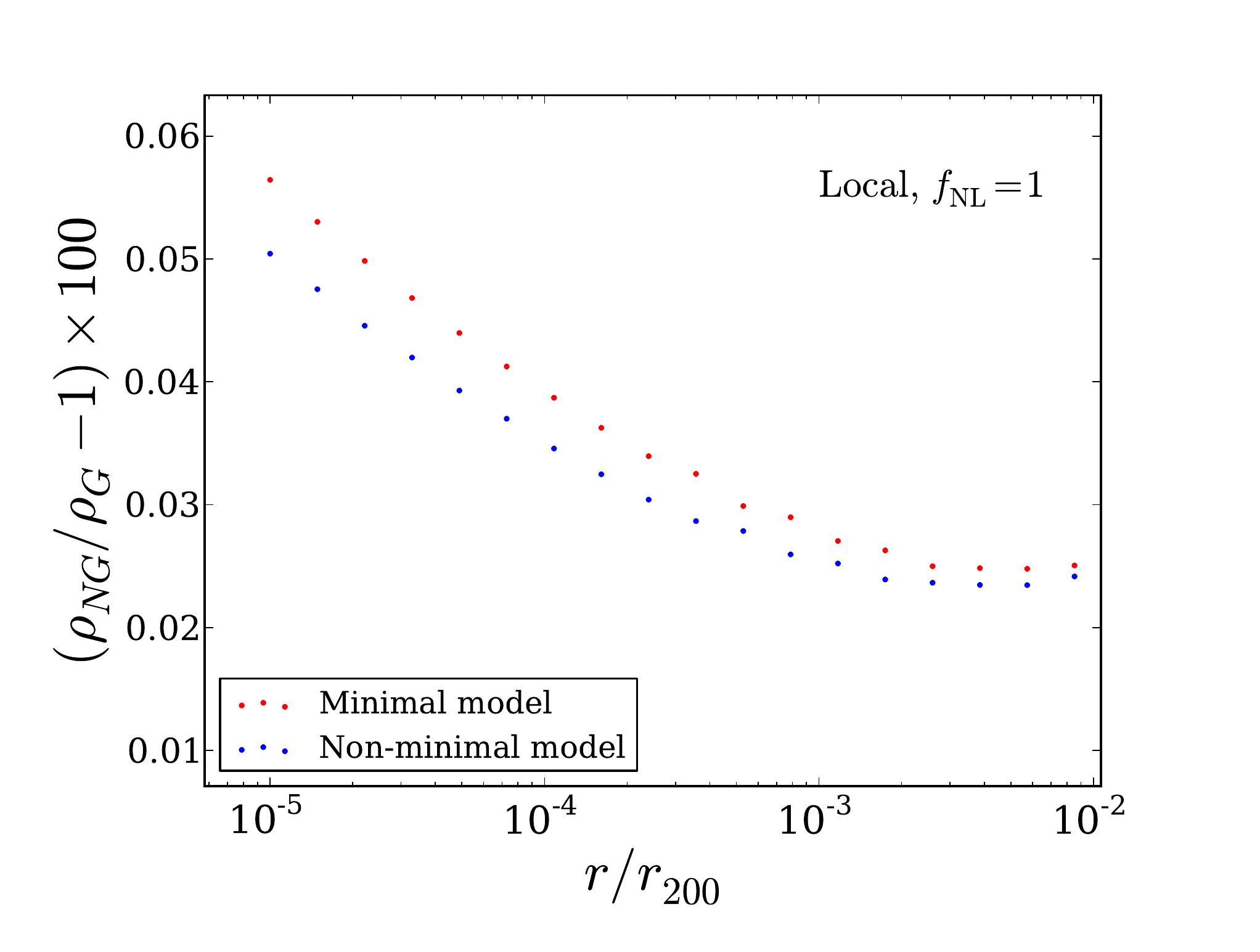}
\includegraphics[width=\columnwidth]{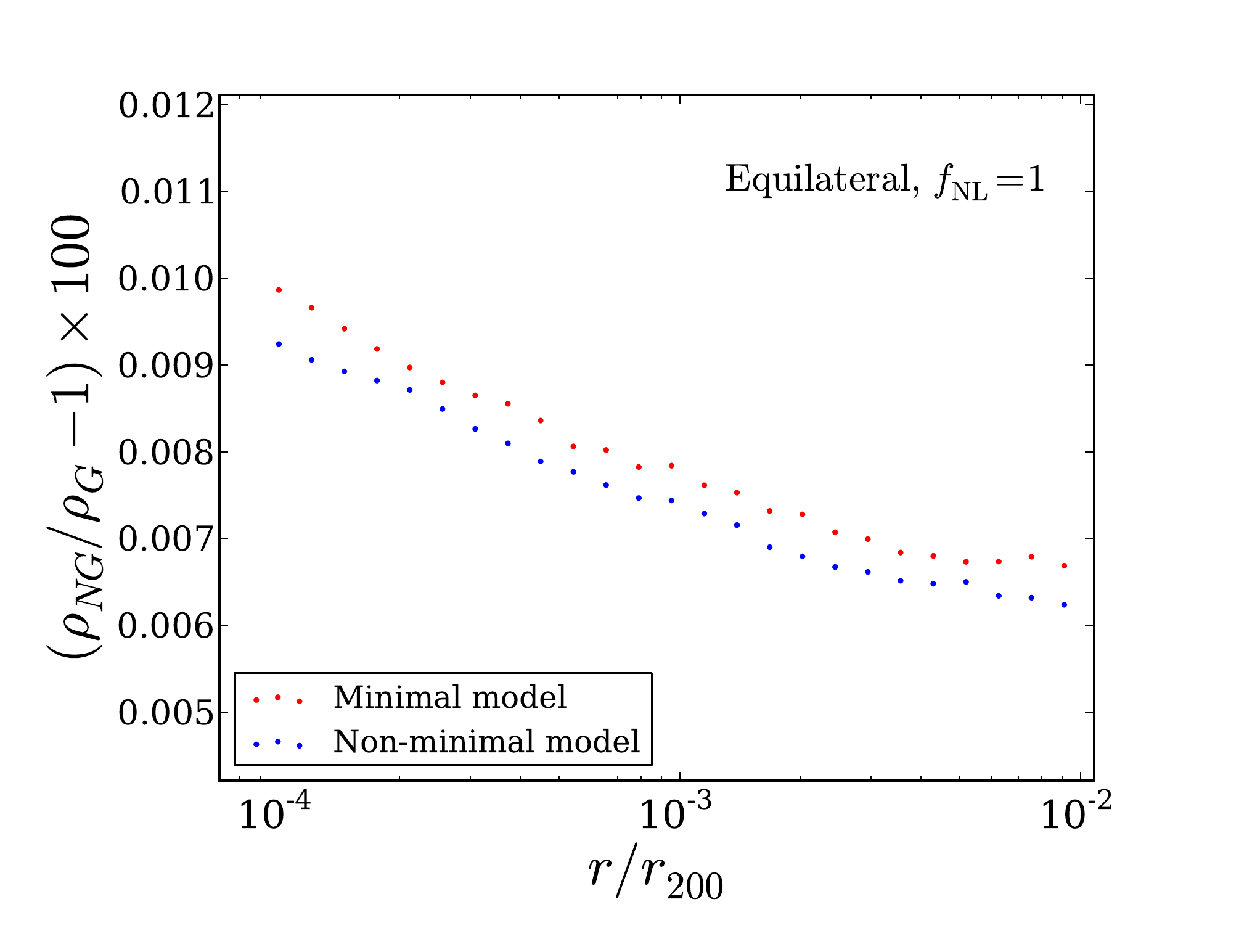}
\caption{The ratio of density profile of the halo for non-Gaussian fluctuations with $f_{\rm NL}=1$ to that of the Gaussian fluctuations. The left panel corresponds to local-shape while the right panel corresponds to equilateral-shape. In both plots, the red curve (the upper curve) shows the ratio in the minimal model of Eq. (\ref{eq:minimal}) while the blue curve (the lower curve) shows the ratio in the nonminimal model of Eq. (\ref{eq:nonminimal}).}
\label{fig:density_ratio}
\end{figure*}
where ${k_3}^2 \equiv {k_1}^2 +{k_2}^2 + 2 k_1 k_2 \mu$. For the local ansatz,
\begin{equation}
\Phi({\bf x})= \phi_{\rm G}({\bf x}) + f_{\rm NL}^{\rm loc} \Bigl[\phi^2_{\rm G}({\bf x})-\langle \phi^2_{\rm G}({\bf x}) \rangle \Bigr],
\end{equation}
the corresponding three-point correlator is  given by 
\begin{equation}
B_{\Phi}^{\rm loc}(k_1,k_2,k_3)=2f_{\rm NL}^{\rm loc}\Bigl[P_{\phi}(k_1)P_{\phi}(k_2) \mbox{ + 2 cyc}.\Bigr],\label{a1}
\end{equation}
where $P_\phi(k)$ is the gravitational potential power spectrum. Inflation models with higher-derivative operators such as the Dirac-Born-Infeld model \cite{Alishahiha:2004eh} give rise to equilateral-shape non-Gaussianity which can be described by the factorizable form \cite{Senatore:2009gt}
\begin{align}
B_{\Phi}^{\rm eq}(k_1,k_2,k_3) &= 6 \ f_{\rm NL}^{\rm eq} \left[-(P_{\phi}(k_1)P_{\phi}(k_2) + 2 {\rm cyc.}) \right. \nonumber  \\
& \left. - 2 {(P_{\phi}(k_1)P_{\phi}(k_2)  P_{\phi}(k_3))}^{2/3} \right. \nonumber \\
& \left. + (P_\phi^{1/3}(k_1)P_\phi^{2/3}(k_2)P_\phi(k_3) + 5 {\rm perm.}) \right].
\end{align}

After evaluating the three-point function, Eq. (\ref{eq:3point}), we calculate the conditional probability given in Eq. (\ref{eq:con_prob_NG}) for non-Gaussian fluctuations for local and equilateral shapes, using fiducial values of $f_{\rm NL}^{\rm loc}= f_{\rm NL}^{\rm eq}= 1$.

Figure \ref{fig:mean_ratio} shows the ratio of the mean of the highest subpeak profile for non-Gaussian fluctuations to that of Gaussian fluctuations. The left panel shows the ratio for the local shape while the right panel corresponds to the equilateral shape. For both local and equilateral shapes, in the inner regions of the halo (small smoothing scale), the mean of the highest subpeak profile is enhanced with respect to the Gaussian case. However, the effect is small, less than $10^{-3}$ even on the smallest scales we have probed. For the equilateral shape the effect is even smaller than in the local case.

Using the model of Ref.~\cite{Dalal:2010hy} to map the peak profile to the mass profile of the collapsed profile, we calculate the non-Gaussian corrections of the local and equilateral shapes to the halo mass and density profiles for the two toy models given in Eq. (\ref{eq:minimal}) and Eq. (\ref{eq:nonminimal}). The ratio of the mass and density profiles for non-Gaussian fluctuations to that for Gaussian fluctuations is plotted in Figs. \ref{fig:mass_ratio} and \ref{fig:density_ratio}. The left panels correspond to the local shape while the right panels correspond to the equilateral shape. For both shapes, the mass and density profiles are enhanced in the inner regions compared to the mass and density profiles for the Gaussian fluctuations, but at a level smaller than $10^{-3}$. Once again the corrections are smaller for the equilateral shape. The results indicate that the non-Gaussian corrections tend to be increasing at smaller radii.

\section{Conclusions}
\label{sec:conclusion}
In this paper we have investigated the influence of primordial non-Gaussianity on the density profile of dark matter halos. Our computation extends the semianalytical model introduced recently by Dalal {\it et al.}~\cite{Dalal:2010hy} which is based on the relation  between the peaks of the initial linear density field and the final density profile of dark matter halos. Our overall conclusion is negative, namely that primordial non-Gaussianity, constrained as it is by recent Planck results~\cite{Ade:2013ydc} ($f_{\rm NL}^{\rm loc} = 2.7 \pm 5.8 \ {\rm and} \ f_{\rm NL}^{\rm eq} = - 42 \pm 75$), is unlikely to have a noticeable effect on halo profiles as might be observed in weak lensing, indirect detection, or large-scale structure.

It is interesting to compare our semianalytic results, based on the formalisms of Refs.~\cite{Dalal:2010hy} and \cite{Maggiore:2009rx}, to the simulations carried out in Ref.~\cite{Smith:2010fh}. Although they simulated more massive halos than the one we have focused on, generally they found corrections to the density profile of order $2-5\%$ for $f_{\rm NL}=100$ (their Fig. 7), which corresponds to $2-5\times 10^{-4}$ for our fiducial $f_{\rm NL}=1$. The simulations of course are limited to relatively large scales, $r\ge 10$ kpc, while our semianalytic approximations are not valid on scales larger than this since the perturbative approach to calculate the non-Gaussian correction breaks at these scales. Nonetheless, Fig. \ref{fig:density_ratio} shows that we find a similar correction to the density profile for the case of local non-Gaussianity. The hint from their Fig. 7 that the effect may be increasing on small scales seems to be borne out by our semianalytic work.

The result that the density increases at inner radii for local and equilateral non-Gaussianity is easily extendable to other shapes of non-Gaussianity. In this sense they might be useful to study the effect of  non-Gaussianity on the matter bispectrum on small scales making use, for instance, of the 
 halo model approach where one needs to  consider non-Gaussian corrections to the halo mass function, the bias functions and  the halo profile \cite{Figueroa:2012ws}. For halo modeling of the two-point function, though, our work seems to suggest that the predictions are insensitive to the changes that primordial non-Gaussianity induces in the halo profile.

\section*{Acknowledgements}
 We thank N. Dalal for useful comments. A.M.D. was funded in part by the U.S. National Science Foundation grant NSF-PHY-1066278. A.R. is supported by the Swiss National Science Foundation (SNSF), project ``The non-Gaussian Universe'' (project number: 200021140236).

\bibliography{SCINC}

\end{document}